# Methodology for assessing system performance loss within a proactive maintenance framework

P. Cocheteux. A. Voisin. E. Levrat. B. Iung

*Centre de Recherche en Automatique de Nancy (CRAN), Nancy-Université, CNRS,
Boulevard des Aiguillettes  B.P. 70239  F-54506 Vandœuvre lès Nancy
(e-mail: {pierre.cocheteux,alexandre.voisin,eric.levrat,benoit.iung}@cran.uhp-nancy.fr)*

Abstract: Maintenance plays now a critical role in manufacturing for achieving important cost savings and competitive advantage while preserving product conditions. It suggests moving from conventional maintenance practices to predictive strategy. Indeed the maintenance action has to be done at the right time based on the system performance and component Remaining Useful Life (RUL) assessed by a prognostic process. In that way, this paper proposes a methodology in order to evaluate the performance loss of the system according to the degradation of component and the deviations of system input flows. This methodology is supported by the neuro-fuzzy tool ANFIS (Adaptive Neuro-Fuzzy Inference Systems) that allows to integrate knowledge from two different sources: expertise and real data. The feasibility and added value of such methodology is then highlighted through an application case extracted from the TELMA platform used for education and research.

## 1. INTRODUCTION

In today's industrial systems, maintenance becomes a lever to reach dependability, safety, quality and user's product requirements (Al-Najjar and Alsyouf, 2003). Maintenance has shift from "fail and fix" practices to become an adding value process for the product and the enterprise. Indeed, besides the traditional "support to production" view, maintenance has a major impact on the performance all along the product life cycle (Takata et al., 2004). In such a view, the aim of maintenance is to ensure the system's aim as well as economical, security… constraints.

According to this new maintenance role, the practices are required to move from corrective maintenance to preventive and even proactive strategies (Djurdanovic et al., 2003). Such maintenance strategy anticipates the failure consequences by evaluating the future evolution of the degradation of the production system. It allows to take the best maintenance decision in relation to criteria evaluated on the future situation of the system, e.g. future production stop, product quality…, to plan the maintenance action (Levrat et al., 2008). The proactivity ability is supported by the prognosis process whose aim is to evaluate the future performances of the system (Muller et al., 2008).

ISO 13381-1 standard defines prognosis as a process for evaluating the Remaining Useful Life (RUL) before failure. In the prognosis domain, some works deal with prediction model for physical state of a component. (Lee et al., 2006) underlines that most of prognostic approach are focused on component prediction as well as the lack of ability to tackle the performance of system.

In order to face these challenges, the paper presents a methodology to build a model for assessing the loss of performance of a system/sub-system/component in the framework of prognosis. The methodology is based on a priori knowledge extracted from functional and dysfunctional point of view of the system.

In this way, this paper is organized in 6 sections. First the context is explained in section 2. The section 3 gives a description of ANFIS. Then the modeling methodology is explained in section 4. Finally an application is presented in section 5 before the conclusion in the last section.

## 2. CONTEXT

### 2.1 Definition of prognosticated performance

The performance can be defined in different way in accordance with the context. In this paper the performance, which has to be prognosticated, is the ability of the system or process to perform its finality. This qualitative notion is supported by quantitative performance indicators. The process finality (goal) is represented by its output flows. Thereby performance indicators are connected with properties of these flows (eg. rotation speed of motor). In order to guarantee system/process performance, performance indicators have to change within a predefined slot of value taking into account the constraint of a minimum maintenance cost.

In this flow-centered vision, the system is described by a succession of processes which consume flows created by upstream processes and which create flows consumed by downstream processes. This functional structure is like a chain of processes and system performance are defined on flows created by the last process. An example of functional description based on TELMA platform (see section 6) is given figure 1.

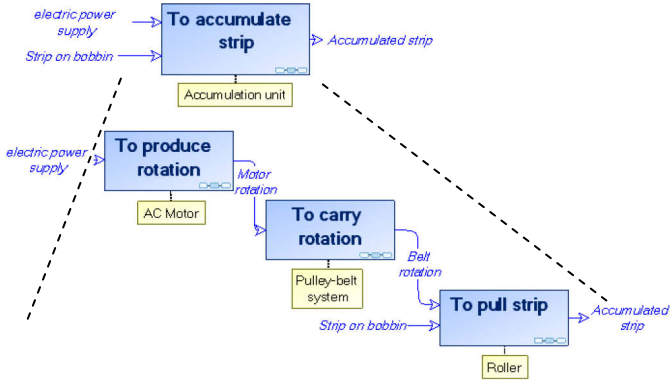

Fig. 1. Process approach of TELMA's accumulation unit

*2.2 Causal Relation of deviation flows*

Performances decrease with time because flow properties evolve and deviate from a nominal value to induce deviation flows. This deviation results from the degradation of support (system or component which support the process) and/or the deviation of input flows. This causality is described by causal relations (Muller et al., 2008) presented in table 1.

**Table 1. Causal relations typology**

|    |         | F   | Flow        | $X_n$ | Nominal state  |
|----|---------|-----|-------------|-------|----------------|
| SP | Support | iF  | Input Flow  | $X_d$ | Degraded state |
| R  | Relation| oF  | Output Flow | $X_f$ | Failed state   |

| Input Flow | Support | Output Flow | Causal relationship | Type |
|---|---|---|---|---|
| $IF_n$ | $SP_n$ | $OF_n$ | $iF_n \wedge SP_n \rightarrow oF_n$ | R1 |
| $IF_d$ | $SP_n$ | $OF_d$ | $iF_d \wedge SP_n \rightarrow oF_d$ | R2 |
| $IF_n$ | $SP_d$ | $OF_d$ | $iF_n \wedge SP_d \rightarrow oF_d$ | R3 |
| $IF_d$ | $SP_d$ | $OF_d$ | $iF_d \wedge SP_d \rightarrow oF_d$ | R4 |
| $IF_d$ | $SP_d$ | $OF_f$ | $iF_d \wedge SP_d \rightarrow oF_f$ | R5 |
| $IF_f$ | $SP_n / SP_d$ or $SP_f$ | $OF_f$ | $iF_f \wedge (SP_n \vee SP_d \vee SP_f) \rightarrow oF_f$ | R6 |
| $IF_n$ or $IF_d$ | $SP_f$ | $OF_f$ | $(Fe_n \vee Fe_d) \wedge SP_f \rightarrow oF_f$ | R7 |

Thus degradation of all components impacts system performances through causal relations. These relations can be classified into three simple models:
- A model of nominal functioning (relation R1),
- A model of the impact of the support degradation on process output flows (relations R3, R4 and R7),
- A model of the impact of the deviation of process input flows on its output flows (relations R2, R4 and R6). This model is also a model of the degradation impact of upstream process supports and of the deviation of system input flows.

*2.3 Prognostic of performance*

In this context, the prognostic of a process performance is performed in two stages:
- the projection of feature evolution (such as degradation indicator) in the future (diachronic view),
- the performance indicator computation which determines the process performance from support degradations and input flow deviations (synchronic view).

Works on prognosis are mainly interested with the projection of a representation of physical state (either a simple variable or a refined indicator which results from several sources) but without interest for performance evolution or state/performance link (Lee et al., 2006).

Thus we propose a methodology which allows to evaluate the performance loss of a system from supports degradation and deviations of system input flows.

This link between degradation and performance is complex and often non-linear. The sources of knowledge about these behaviours are mainly (human) expertise and/or experience data and give, most of time, an incomplete and qualitative knowledge. Expert knowledge is formulated in a linguistic and qualitative way while data represent a limited number of behaviour. A neuro-fuzzy tool can model this link by integrating the linguistic knowledge and by using knowledge from data to refine the model.

ANFIS, explained in the next section, can support this methodology and the view about a process is shown in figure 2.

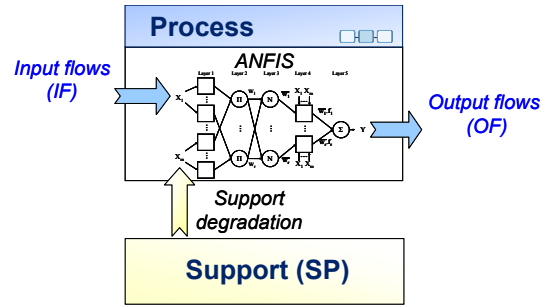

Fig. 2. Performance model of a process

### 3. ADAPTATIVE NEURO-FUZZY INFERENCE SYSTEMS (ANFIS)

The use of fuzzy neural networks (FNN) in proactive maintenance framework is recent. ANFIS is a FNN proposed by (Jang, 1993) and is the most common one. It is used to predict time series (El-Koujok et al., 2008).

In this paper, ANFIS is used to model the performance degradation. On the contrary of classical neural networks, it allows to integrate dysfunctional knowledge by using fuzzy rules and to keep learning ability of neural network to adapt from data.

The model has 5 layers. Figure 3 presents an ANFIS with *m* input $X_i$.

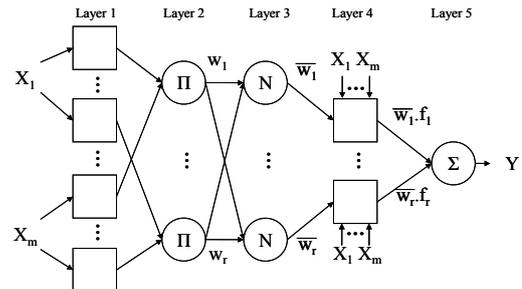

Fig. 3: ANFIS structure

We use $S_{ij}$ for the output of the $j^{th}$ neron of the $i^{th}$ layer.

- 1$^{st}$ layer : Fuzzyfication

This layer is dedicated to the conversion of the input variables into the fuzzy descriptor according to fuzzy sets.

$$S_{1j} = \mu_{lk}(X_k) \quad avec \quad l=1,...,nk \ et \ k=1,...,m \quad (1)$$

where $\mu_{lk}(X_k)$ is the membership degree of the $k^{th}$ input variable to the $l^{th}$ term of the fuzzy partition. The parameters which define these functions are called premise parameters.

- 2$^{nd}$ layer : Fuzzy rules weighting

Each neuron corresponds to a fuzzy rule. The value computed by the neuron equals to the weight of the fuzzy rule regarding its inputs and using a T-norm. The most common T-norm used is the product:

$$S_{2,j} = \prod \mu_{lk} = w_j \quad (2)$$

where $\mu_{lk}$ is the membership degree of $X_k$ to the $lj^{th}$ linguistic terms used in the $j^{th}$ rule.

- 3$^{rd}$ layer : Normalisation

The aim of this layer is to normalize weights.

$$S_{3,j} = \frac{w_j}{\sum_{k=1}^{k=r} w_k} = \overline{w_j} \quad (3)$$

for j = 1,…, r.

- 4$^{th}$ layer : Defuzzyfication

The output is a linear combination of the input values weighted by the rule as defined in Takagi-Sugeno rules.

$$S_{4,j} = \overline{w_j}.f_j \quad avec \quad f_j = \sum_k a_{jk}.X_k + b_j \quad où \ k=1,..,r \quad (4)$$

$a_{jk}$ et $b_j$ parameters of $f_j$ are called consequent parameters.

- 5$^{th}$ layer : Output computation

The final output Y is computed with:

$$S_5 = \sum_{j=1}^{j=r} S_{4,j} = \sum_{j=1}^{j=r} \overline{w_j}.f_j \quad (5)$$

The ANFIS parameters are computed during a supervised training phase. Thus ANFIS needs data which associate input sequences and outputs. This training is performed in a recursive way in order to reach an optimum parameter. The used hybrid training algorithm is explained in (Jang, 1993). It is a combination of the gradient descent approach and the least square method. The first one tunes premises parameters by fixing consequent parameters and the second tunes the consequent parameters by fixing premises parameters. The learning continues until the desired number of training step (epochs) or the requirement for stopping is reached.

## 4. MODELING METHODOLOGY

The methodology purpose is to propose design way of a model which evaluates system performance indicators by integration of knowledge from expertise and data. The neuro-fuzzy tool ANFIS is used to performed this integration.

### 4.1 Causal relations instantiation

In a first step, causal relations have to be defined by instancing generic relations according to expert knowledge. The functional and dysfunctional knowledge can be modelled with tools like process approach which allows to describe the functional structure of a hierarchical multi-level system. The dysfunctional knowledge is extracted from FMECA (Failure Mode, Effects, and Criticality Analysis) and HAZOP (HAZard and OPerability) studies. HAZOP study allows for each flow of a system to connect deviations with their causes and consequences. The standard IEC 61882 proposes generic deviations (table 2).

**Table 2. Generic deviations**

| NO | Complete negation |
|---|---|
| MORE | Quantitative increase |
| LESS | Quantitative decrease |
| PART OF | Qualitative decrease |
| AS WELL AS | Qualitative increase |
| OTHER THAN | Complete substitution |
| REVERSE | Logical opposite |

In this paper the considered properties are defined on more specific flow of energy or material and are only quantitative. The set of generic deviations can be limited to {NO, LESS, MORE}.

**Table 3. Part of HAZOP study**

| | Deviations | Causes |
|---|---|---|
| Output | NO (output) | NO (input) |
| | | Failure mode FM1 |
| | LESS (output) | LESS (input) |
| | | Failure Mode FM2 |

**Table 4. Causal relations**

| Relation | Type |
|---|---|
| OK (input) ∧ Healthy → OK (output) | R1 |
| LESS (input) ∧ Healthy → LESS (output) | R2 |
| OK (input) ∧ FM2 → LESS (output) | R3 |
| LESS (input) ∧ FM2 → LESS (output) | R4 |
| NO (input) ∧ (Healthy ∨ FM2 ∨ FM1) → NO (output) | R6 |
| (OK (input) ∨ LESS (input)) ∧ (FM1) → NO (output) | R7 |

The instantiation of causal relations using knowledge from HAZOP study leads to a set of relation which represents the impact of the deviation of input flows and degradation/deterioration of supports. In table 1, a flow is described using three states: nominal, degraded and failed. The degraded state represents a partial performance loss and is connected with MORE or LESS deviations. The failed state means a complete performance loss and corresponds to the NO deviation. The deviation of nominal state is zero and the property is OK.
The causal relations are instanced with the HAZOP study presented table 3. The resulting causal relations are presented in table 4. The healthy state corresponds to a support without current failure mode.

*4.2 Definition of fuzzy rules set and integration in ANFIS*

The second step is the translation of the causal relations previously obtained into a set of fuzzy rules which can be introduced in the ANFIS. One ANFIS structural constraint is that one rule is required for each output fuzzy set. The relations of table 4 are re-written into fuzzy rules (table 5).

**Table 5. Obtained rules set.**

| Rules | Type |
|---|---|
| OK (input) ∧ Healthy → OK (output) | R1 |
| LESS (input) ∨ FM2 → LESS (output) | R2, R3 & R4 |
| NO (input) ∨ (FM1) → NO (output) | R6 & R7 |

In order to define the FIS (Fuzzy Inference System) inputs membership functions (MF) has to be defined. Using a strict partition, premises parameters can be obtained from expert information like MF shape (triangular, trapezoid, gaussian…) and/or MF significant parameters.

*4.3 Training of consequent parameters*

First and second step define rules and premises parameters of ANFIS. In the third step, the consequent parameters are computed in a training phase. Data are used to perform a supervised training. The hybrid training algorithm is limited to the least square method.

*4.4 System performance model*

The three first steps allow to obtain a tool which compute the impact of input deviation and component degradation/deterioration on output deviation of this component. In the last step, the process approach is used to reach the system level from the component level. The ANFIS output of an upstream process is connected with the ANFIS input of the next process. Thus the inputs of the system performance model are the inputs of first processes and the components degradations/deteriorations levels. Its outputs are the outputs of last processes. The figure 4 shows the system performance model of the application case used in the next section.

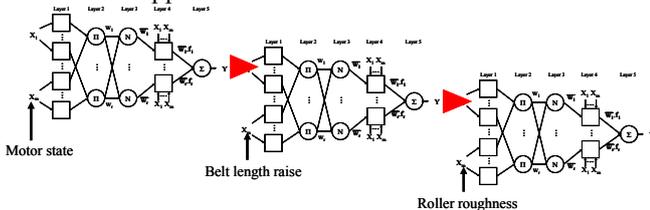

Fig. 4: System performance model

## 5. APPLICATION AND RESULTS

The feasibility and interest of the method are considered in relation to an application case extracted from TELMA platform. TELMA (figure 5) is a laboratory testing platform materializing a physical process dedicated to unwinding and stamping metal strip (Levrat and Iung, 2007). This process is similar to concrete industrial applications such as sheet metal cutting and paper bobbin cutting. The physical process is divided into four parts: bobbin changing, strip accumulation, punching-cutting and advance system. The application case focuses on the accumulation unit. This application is supported by the Matlab© "Fuzzy Logic" toolbox.

*5.1 Application on accumulation unit*

The accumulation unit is composed of an AC motor, a pulley-belt system and a roller. Its functional analysis is given figure 1.

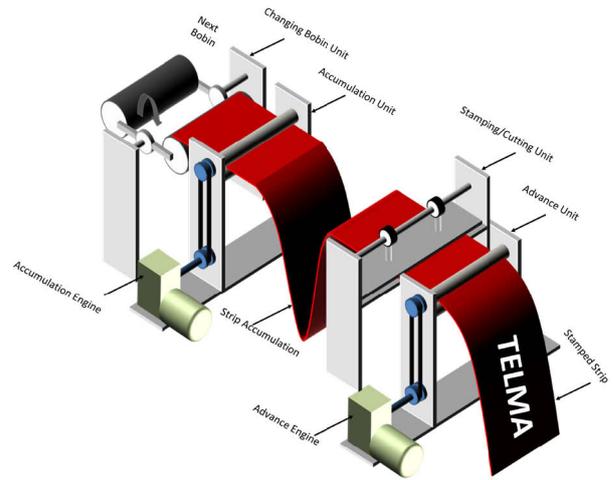

Fig. 5 : TELMA Platform

AMDEC and HAZOP studies were made on the platform. Table 6 present the part of the HAZOP study dealing with output flows of the three sub-functions which compose "to accumulate strip". The chosen property of the rotation flow is angular velocity (Wom for the motor output and Wob for the belt one) and the one related to the "accumulated strip' flow is the average strip flow rate (QStrip). In the following we suppose that there is always some strip provided by the precedent unit (no deviation of "strip on bobbin"), i.e. there is always a bobbin with strip on the changing bobbin unit.

The AC motor has several degradation modes due to different stresses (Bonnett, 2000). In this application, we only consider failure modes of the AC motor because the impact of degradation modes on rotation speed is neglected. Thus its state can be OK or failed. The pulley-belt system is composed of elements which have different degradation dynamics. We only consider the degradation of the belt which is faster than the pulley degradation. The main belt degradation is the lengthening. The lengthening causes a degradation (lengthening) until the failure (slipping). Finally the roughness diminution of the surface's roller reduces the friction between strip and roller and creates an intermittent slip (degradation) until the failure (full slip).

With respect to the method detailed in section 4, fuzzy rules are built. The example of the belt is presented table 7.

**Table 6. HAZOP studies of accumulation unit**

| AC motor rotation | | |
|---|---|---|
| Attribute | Deviation | Cause |
| Angular Velocity (Wom) | NO | NO current (electric power supply) |
| | | Failure (AC motor) |
| | LESS | LESS current (electric power supply) |
| | MORE | MORE current (electric power supply) |
| **Belt rotation** | | |
| Attribute | Deviation | Cause |
| Angular Velocity (Wob) | NO | NO Angular Velocity (AC motor rotation) |
| | | Slip (belt) |
| | LESS | LESS Angular Velocity (AC motor rotation) |
| | | lengthening (belt) |
| | MORE | MORE Angular Velocity (AC motor rotation) |
| **Accumulated strip** | | |
| Attribute | Deviation | Cause |
| Average strip flow rate (QStrip) | NO | NO Angular Velocity (Belt rotation) |
| | | Slip (roller) |
| | LESS | LESS Angular Velocity (Belt rotation) |
| | | Intermittent slip (roller) |
| | MORE | MORE Angular Velocity (Belt rotation) |

**Table 7. Fuzzy rules of the belt**

| |
|---|
| OK (Wom) ∧ Healthy (belt) → OK (Wob) |
| LESS (Wom) ∨ lengthening (belt) → LESS (Wob) |
| NO (Wom) ∨ Slip (belt) → NO (Wob) |
| MORE (Wom) ∧ Healthy (belt) → MORE (Wob) |

The input MFs of the ANFIS are given by experts. Deviations of flows are defined with respect to their nominal values and are given as percentage of the nominal value. This expertise is presented in table 8.

**Table 8. Expertise result on flow property**

| OK (nominal value) | 100% |
|---|---|
| NO | 0% |
| LESS | Between 0 and 100% |
| MORE | Upper than 100% |

As only few part of knowledge about the flow's MFs are available, linear functions are chosen (i.e. triangular and trapezoid function). With the constraint of strict partition, the MFs can be defined and is shown in figure 6. Such a partition has been chosen since we expect gradual output from 0% to 100% and the consequent part of the rule is constant. Hence the gradual behaviour comes from the triangular MF.

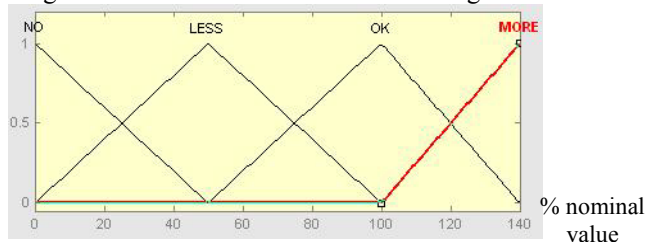

Fig. 6: Flow's MFs

The result of expertise about the belt's lengthening (define by the length raise) and the roller's roughness (define by the roughness average Ra) are given by table 9 and allow to define the MFs (figure 7a and 7b).

**Table 9. Expertise result**

| | | |
|---|---|---|
| Belt's length raise | OK (nominal value) | Between 0 and 0.5 mm |
| | Deg (lengthening) | Average value = 4.5 mm |
| | Fail (Slip) | Upper than 8.5 mm |
| Roller's roughness (Ra) | OK (nominal value) | Upper than 12.5 μm |
| | Deg (intermittent slip) | Average value = 7.85 μm |
| | Fail (Slip) | Lower than 3.2 μm |

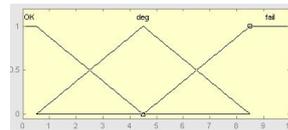 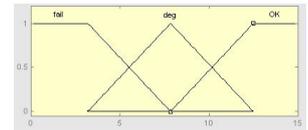

Fig. 7a: Lengthening MFs    Fig. 6b: Ra's MFs

Finally motor's MFs have to be defined. The motor state is a discrete variable and only continuous variables are accepted by Matlab fuzzy logic toolbox. Thus a continuous state variable S, defined on the [0,1] interval, has to be created with constrains that:
- if motor's state is OK then $\mu_{OK}(S) = 1$ and $\mu_{failed}(S) = 0$,
- if motor's state is Failed then $\mu_{OK}(S) = 0$ and $\mu_{failed}(S) = 1$,

with $\mu_{OK}(S)$ and $\mu_{failed}(S)$ the membership degree of S to the OK and Failed terms. MFs behaviour between 0 and 1 is not considered since the S equals either 0 or 1.

The consequent parameters of the ANFIS are computed during the training phase using the data sets. These trainings have been performed on 500 epochs. In this application, the consequent parameters are chosen as constant (input's coefficients are zero).

*5.2 Results*

The objective of this experimentation is to validate the models' behaviors and to visualize the component degradation impacts on the system output flows. Two scenarios have been proposed:
(1) the input flow and the motor have no deviation/degradation while belt and roller are degrading. The belt degradation dynamic is higher than the roller one. Thus a maintenance action (overhaul) is performed on the belt at the middle of scenario, i.e. the belt degradation is reset at this time.
(2) a failure of the motor occurs without deviation/degradation.

The figure 8 shows results of the first scenario. The evolution of the degradation of the belt (length raise) and the roller (Ra) are presented as well as the average strip flow rate (QStrip). Both degradation processes have two different dynamics of degradation: a faster one (belt) and a slower one (roller). The QStrip curve shows the deviation of the system performance and shows the impact of both degradation processes on it.

During the maintenance action, the process is stopped, i.e. the degradation of the roller is stopped too, and an overhaul of the belt is performed. After the maintenance action, the degradation of the belt is reset (length raise = 0mm). The consequence on QStrip is an enhancement but not a reset of the deviation (QStrip = 96.15%). This is due to the degradation of the roughness which is not impacted by the maintenance action. Thus QStrip, for an identical length raise, is higher before the maintenance action. This observation points up that the proposed model allows to consider and quantify the impact of degradations from different components on the system performance.

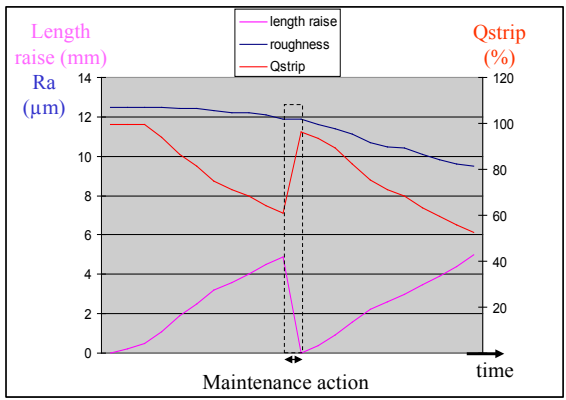

Fig. 8: Results of the first scenario

Figure 9 shows results of the second scenario. The evolution of rotations of the motor (Wom) and belt (Wob) and the average strip flow rate (QStrip) are presented as well as the motor's state. This scenario allows to show the influence of the motor's failure and its propagation on performances of downstream functions through their input flows. Indeed when the motor fail, the motor rotation speed becomes null. Thus, because of causal relations, the property of output flows of the downstream functions becomes null as well and propagate the failure through the system.

A second observation can be made: the level of Wom, Wob and Qstrip aren't exactly equal while it should be (in this case 100% before failure, 0% after it). This is due to the quality of the training phase and data. It points out that the training phase must not be neglected and the method needs quantity and quality of data.

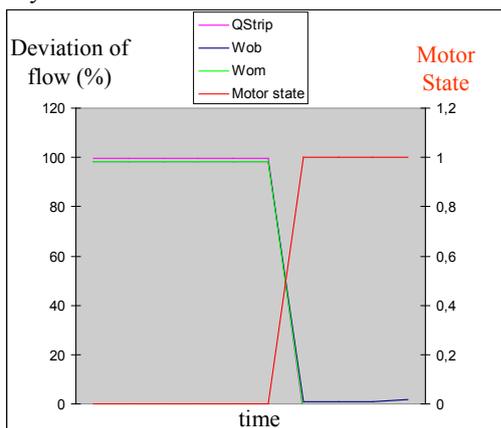

Fig. 9: Results of the second scenario

6. CONCLUSIONS

This paper proposes a methodology that allows the evaluation of system performance loss. This methodology is supported by the neuro-fuzzy tool ANFIS. This tool allows to integrate expert's knowledge and to extract knowledge from data. The expert knowledge is combined with causal relations in order to obtained a fuzzy rules set. This first model of the link degradation/performance is simple and rough but sufficient to built the ANFIS architecture. The knowledge from data is used in a training phase to obtain the ANFIS consequent parameters.

An application case has been presented. It shows building of the model according to expert's knowledge and data training. Then, two scenarios are discussed in order to show the impact of the component degradation/failure on the system performance.

The combination of knowledge coming from different sources allows to reduce drawbacks of a single knowledge approach: the expertise reduces the lack of sufficiency of data and data refine the expert knowledge.

In future work, prognostic models of degradation have to be plugged in order to obtain a complete prognostic model. The use of stochastic prognostic model of degradation with this system performance model will be study.